\documentclass[aps,prl,twocolumn,groupedaddress,amssymb]{revtex4}
\usepackage{graphicx,latexsym}

\begin{document}
\def\be{\begin{equation}}
\def\ee{\end{equation}}
\def\ba{\begin{eqnarray}}
\def\ea{\end{eqnarray}}
\title{Seeing Anderson Localization}


\author{M. Hilke}



\affiliation{Dpt. of Physics, McGill University, Montr\'eal, Canada
H3A 2T8}

\begin{abstract}
Anderson localization was discovered 50 years ago to describe the
propagation of electrons in the presence of disorder
\cite{Anderson58}. The main prediction back then, was the existence of disorder
induced localized states, which do not conduct electricity. Many years later it turns
out, that the concept of Anderson localization is much more general
and applies to almost any type of propagation in time or space,
when more than one parameter is relevant (like phase and
amplitude). Here we propose a new optical scheme
to literally {\em see} Anderson localization by varying the
optical wavelength or angle of incidence to tune between localized and
delocalized states. The occurrence of Anderson localization
in the propagation of light, in particular, has become the focus of tremendous interest due to the
emergence of new optical technologies and media, such as low dimensional 
and disordered optical lattices \cite{Schwartz07,Lahini08}. While several experiments have reported the measurement of Anderson localization of light \cite{Wiersma97,Berry97,Storzer06,Schwartz07,Lahini08}, many of the
observations remain controversial because the effects of absorption
and localization have a similar signature, i.e., exponential decrease
of the transmission with the system size
\cite{Scheffold99,Chabanov00}. In this work, we discuss a system, where we
can clearly differentiate between absorption and
localization effects because this system is equivalent to a {\em perfect} filter, only in
the absence of any absorption. Indeed, only one wavelength is perfectly
transmitted and all others are fully localized. These
results were obtained by developing a new theoretical framework for the average
optical transmission through disordered media.
\end{abstract}

\maketitle

In order to literally {\em see} Anderson localization, we consider the
system shown in figure 1. The setup is composed of $N_f$ filters in
series. Each filter is composed of $N_l$ random optical layers and each
layer has a refractive index $n_j$ and a thickness $d_j$ as defined
in figure 2. For optimal results, it is important that these layers are very well
defined. This is possible, for instance, by using high accuracy
multilayer growth techniques, such as molecular beam epitaxy (MBE).
Good material choices include large band gap materials like GaN ($\Delta=3.2eV$,
$n\simeq 2.3$) and InN ($\Delta=2eV$, $n\simeq 3.1$),
which can be grown by MBE with atomic precision \cite{InN,InN2}. The absence of any surface
roughness ensures that this multilayered system is one-dimensional
in nature. Surface roughness would induce scattering at stray angles,
which leads to a 1D to 3D dimensional crossover
\cite{Milner08}, thereby strongly reducing localization effects,
since localization in 3D is much weaker than in 1D \cite{Abrahams79}
and therefore much more challenging to observe \cite{Storzer06}.

\begin{figure}
[ptb]
\includegraphics[width=3in]{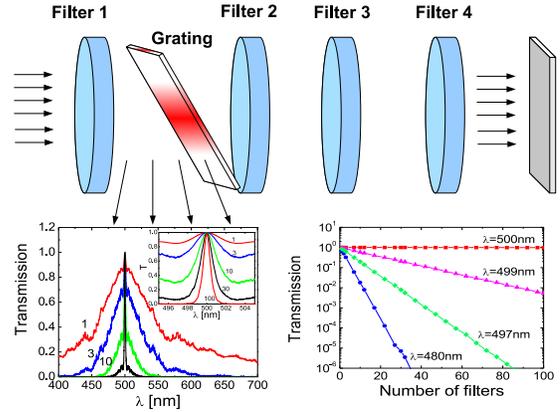}\\%
\caption{The top shows the arrangement of the filters, where a
grating can be inserted in order to measure the wavelength dependence
of the transmission. This transmission is shown in the bottom left figure when
the grating is placed after 1, 3, 10, 30, or 100 filters, which increases the quality factor filter with
the number of layers. The bottom right figure represents the
dependence of the transmission on the number of layers at several wavelengths
(480, 497, 499, and 500)nm, indicating an exponential decay}%
\label{3D}%
\end{figure}

To illustrate the general behavior of our system, we considered the
simplest random system, a binary distribution, where we assume two kinds of
materials, one with a refraction index of $n_0=1$ and the other with
$n_1=3$. Adding more materials will not change the results
qualitatively and the analytical expressions derived below remain
valid for any choice or combination of materials. However, to maximize the
effect of disorder (smallest localization
length) the variance of refraction indices should be largest. Since in
most materials the refraction index rarely exceeds $n=3$, an
efficient way to increase the variance is to consider a discrete
distribution. Indeed, for a binary random distribution \{$n_0=1$ and
$n_1=3$\} the variance is one as compared to a uniform random
distribution $1\leq n \leq 3$, where the variance is reduced to one-third.

\begin{figure}
[ptb]
\includegraphics[width=3in]{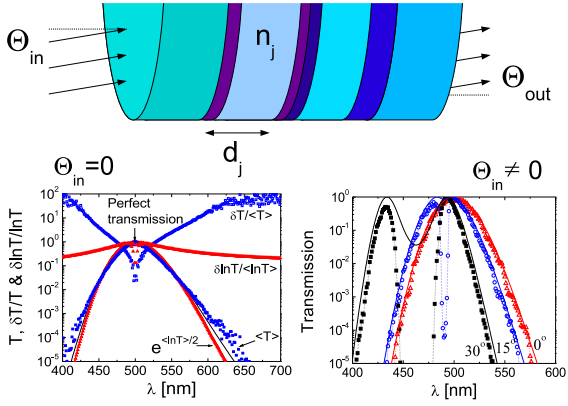}\\%
\caption{The top picture illustrates the details of a random sequence of
layers composing one filter. In the bottom left figure the transmission as a
function of wavelength is shown for a system of 200 layers averaged
$10^4$ times. The analytical result from equation (\ref{analytic}) is shown with a continuous line.
The numerical average and relative fluctuations are shown with square and triangles.
The transmission as a function of wavelength for
different incidence angles (0, 15 and 30$^\circ$) is depicted in the bottom right figure, where the continuous lines
correspond to the analytical expression (\ref{analytic}), whereas the dotted lines correspond to the
numerical log-average.}%
\label{3D}%
\end{figure}

The filters are placed in a way to allow the insertion of a grating between two of them. The
grating is necessary to resolve the spectral composition of the transmitted light.
The position of the grating can be changed in order to measure the
transmission after 1, 2,\ldots, or $N_f$ filters in series. When using a white light source,
this allows to spectrally resolve the transmission after any number of filters.
Anderson localization will then lead to an exponential decay of the transmission with the
number of filters (see figure 1).

A major difficulty in observing Anderson localization is the
existence of strong fluctuations \cite{Chabanov00,Genack05}. In general, the relative
fluctuations of the transmission $(\delta T/\langle T\rangle)$
diverge with decreasing transmission, where $\delta
T=\sqrt{(T-\langle T\rangle)^2}$ and $\langle T\rangle$ is the
average transmission. This is a fundamental problem, which can be
circumvented by considering $\ln(T)$ (the transmission in decibel)
or the inverse transmission since in both cases the
relative fluctuations decrease with decreasing transmission as shown
in figure 2 for $\ln (T)$. Therefore, in order to reduce the fluctuations we
can average the logarithm of the
transmission over several disorder configurations, which converges much
faster than an average over the transmission. The numerical examples
shown in figure 1 are based on a small log-average over 100
configurations and assuming 100 different filters, each with an
average of 50 random layers and a spectral resolution of 1nm. Such
a log-average over a small number of configurations is feasible
experimentally, since it simply involves averaging the signal in decibel
over a few configurations. In comparison, a higher averaging is used
in figure 2, where $10^4$ configurations
with 200 random layers was used, leading to even less fluctuations. Considering a system of $N_f$ different filters allows one to perform a configurational average over the disorder, where the space between two filters is equivalent to an additional layer, as long as the distance between two filters follows the resonance condition described below. However, this is equivalent to simply considering a single filter with more random layers and then performing a configurational average, which is what we use in the remainder of this letter.

The results were obtained using a standard method to
describe a multilayered optical system with a transfer matrix. Each
matrix describes the transmission after one layer, where we assume
the layers to be normal to the $\hat{x}$ direction. For TM
(transverse magnetic) waves, where the magnetic field component
$H^z$ is parallel to each layer and the electric field component
$E^y$ is at an angle $\theta$ (the angle of incidence) to the first
layer, the transfer matrix can be written as \cite{Chilwell84}:

\be \left(\begin{array}{c}
H^z_{j-1}\\E^y_{j-1}\end{array}\right)=\underbrace{\left(\begin{array}{cc} \cos(\phi_j) &
-i\sin(\phi_j)/\gamma_j\\-i\sin(\phi_j)\gamma_j & \cos(\phi_j)
\end{array}\right)}_{M_j}\left(\begin{array}{c}
H^z_{j}\\E^y_{j}\end{array}\right), \label{iter1}\ee

For an incoming wave, of wavelength $\lambda$, the field components on one side of a stack
of $N_l$ layers can therefore be related to the fields on the other side
by taking the product of the transfer matrices
($M=\prod_{j=1}^{N_l}M_j$). Assuming a vacuum impedance $z_0$ before
and after the filter, the transmission of light is given by
$T=4|M_{11}+\gamma M_{12}+M_{21}/\gamma+M_{22}|^{-2}$, where
$M_{kl}$ are the matrix elements of $M$ \cite{Chilwell84}. The
matrix elements of $M_j$ used in equation (\ref{iter1}) are material
dependent with $\phi_j=(2\pi
d_j/\lambda)\sqrt{n_j^2-\sin^2(\theta)}$ and
$\gamma_j=(z_0/n_j^2)\sqrt{n_j^2-\sin^2(\theta)}$. For TE
(transverse electrical) waves these expressions have to be replaced
by $H^z\rightarrow E^z$, $E^y\rightarrow -H^y$, and
$\gamma_j=z_0^{-1}\sqrt{n_j^2-\sin^2(\theta)}$.

Absorption exists when the imaginary part of the refraction
index is non-zero ($\Im\{n_j\}\neq 0$). $\Im\{n_j\}$ can be substantial
when photon energies approach the band gap energy or when the materials are
conducting. With absorption the localization-delocalization transition, described below,
cannot be observed. For instance, for 100 layers and an imaginary part of the refraction index of only 0.01 the
transmission at the resonance is decreased by about a factor 100 so that absorption effects can unambiguously be
distinguished from localization effects. This is crucial, and in the remainder of this
letter we will only consider the case without absorption, for which the localization-delocalization transition
can be obtained under the following condition:

For a given layer there exists a set of thicknesses
$d_j=m\lambda_0/2n_j$ ($m$ any integer) for which the layer is
totally transparent at normal incidence and for the resonant
wavelength $\lambda_0$. This corresponds to $\sin(\phi_j)=0$ and to
$M_j$ equal to the identity matrix. When randomly combining different layers
with the same resonance condition, the overall resonance
condition is preserved. The position of these resonances was discussed by
Cristanti \cite{Crisanti90}. Away from the resonant wavelength the
system behaves like a random system, leading to Anderson
localization. This type of disorder was also studied in the context of
electronic transport in 1D and 2D \cite{corr1,corr2,corr3}, where metal-insulator
type transitions were found. Figures 1 and 2 show this resonance,
where the transmission is one (transparent) at $\lambda_0=500$nm and then
decreases exponentially ($T\simeq e^{-K(\lambda-\lambda_0)^2}$) away
from the resonance condition. $K$ is a form factor which is proportional to the number of random layers and
also depends on the variance of the disorder as discussed below. This dependence on the number of layers can then
be used to tune the quality factor of the filter. Without this special resonance condition, the
transmission through random optical layers decays exponentially
at all wavelengths \cite{Bouchaud86,Berry97,Bliokh04}.

We now turn to characterize the transmission through this random
system by evaluating the product of $M_j$'s composed of
random elements. Similar products of random matrices have been
widely studied and lead to matrix elements, which increase exponentially with the number of products.
The rate of this exponential dependence is termed the
Lyapunov exponent and will be derived analytically below. The Lyapunov exponent also
corresponds to the inverse of the localization length, beyond which the transmission
vanishes. To obtain analytical results the trick is to consider the square of the fields,
$|H^z_j|^2$ instead of $H^z_j$. This allows to concentrate on the
dominant behavior beyond the plane wave oscillations. A similar technique
was actually used in the context of electronic transport
\cite{Erdos89,pendry94,Hilke08} and we extend it here to optical
systems. Squaring equation (\ref{iter1}) then leads to the following
iterative equation:

\begin{widetext}

\be \underbrace{\left(\begin{array}{c} |H^z_{j-1}|^2\\|E^y_{j-1}|^2\\2\Im
\{H^z_{j-1}(E^{y}_{j-1})^*\}\end{array}\right)}_{\vec{F}_{j-1}}=\underbrace{
\left(\begin{array}{ccc} \cos^2(\phi_j) & (\sin(\phi_j)/\gamma_j)^2
& i\cos(\phi_j)\sin(\phi_j)/\gamma_j \\ (\sin(\phi_j)\gamma_j)^2 &
\cos^2(\phi_j) & -i\cos(\phi_j)\sin(\phi_j)\gamma_j\\ 0 & 0 &
\cos^2(\phi_j)-\sin^2(\phi_j)
\end{array}\right)}_{M^{3\times
3}_{j}} \underbrace{\left(\begin{array}{c}
|H^z_{j}|^2\\|E^y_{j}|^2\\2\Im\{(H^z_{j}(E^{y}_{j})^*\}\end{array}\right)}_{\vec{F}_{j}}.\ee
\end{widetext}

For $N_l$ layers, the total transmission is now determined by the
product $ \vec{F_0}=(\prod_{j=1}^{N_l}M^{3\times
3}_{j})\vec{F}_{N_l}$, which can be averaged over the disorder,
leading to, $\langle\vec{F_0}\rangle=\langle M^{3\times
3}\rangle^{N_l}\langle\vec{F_{N_l}}\rangle$, where $\langle
M^{3\times 3}\rangle$ is the disorder average of $M^{3\times 3}_{j}$
assuming that the material parameters of the layers are not
correlated. The disorder average depends on the distribution of the
parameters entering $M^{3\times 3}_{j}$. The leading behavior is
obtained by taking the eigenvalues of $\langle M^{3\times
3}\rangle$, from which the Lyapunov exponent can be extracted as
$\Lambda=\max\{\ln|\mbox{Eig}(\langle M^{3\times 3}\rangle|)\}$.
This yields an expression for the average resistance (inverse
transmission), where $\langle 1/T\rangle=e^{\Lambda N_l}$. Hence, perfect
transmission corresponds to the case where $\Lambda=0$. Interestingly, this
expression can also be related to the log-average
transmission $\langle\ln T\rangle= -\Lambda N_l/2$ and to the
average transmission $\langle T\rangle=e^{-\Lambda N_l/4}$, where
the factors of 1/2 stem from the properties of the distribution of
$T$'s \cite{pendry94}. The next step is obtaining $\langle
M^{3\times 3}\rangle$ for a given distribution. For discrete
distributions, like the binary one this is quite straightforward,
since the layer properties are determined by only two possible configurations,
\{$d_0$, $n_0$\} and \{$d_1$, $n_1$\}, respectively. This yields

\be \langle M^{3\times 3}\rangle=(M^{3\times 3}_0+M^{3\times
3}_1)/2, \label{analytic}\ee

where $M^{3\times 3}_0$ and $M^{3\times 3}_1$ are simply $M^{3\times
3}_j$ with \{$d_j=d_0$, $n_j=n_0$ \} and \{$d_j=d_1$, $n_j=n_1$ \},
respectively. For continuous distributions, analytical expressions
for $\langle M^{3\times 3}\rangle$ are often too complicated to be
useful and can be integrated numerically instead.

Comparing these analytical expressions to the numerical results demonstrate
a remarkable agreement as seen in figures 1 and 2. The analytical
expression using equation (\ref{analytic}) reproduces all the main
features and shows the resonance behavior at $\lambda=\lambda_0$.
Evaluating the logarithm of the eigenvalues of expression (\ref{analytic}) close to
the resonance gives $\Lambda\sim(\lambda-\lambda_0)^2$, which is
perfectly reproduced by our numerical results and determines the form factor $K$, which depends on the variance of the disorder. (Zero variance leads to $K=0$ or $\Lambda=0$).
Deviations exist when looking at the angle dependence, as shown in figure 2. While the
analytical expression predicts two transmission peaks with a
weak minimum in between, the numerical results show that the minimum
is more pronounced. This can be attributed to restricting the average to the second
moment, which can lead to differences with the numerical
results and in some cases even lead to fluctuations in the Lyapunov
exponent \cite{Hilke08}.

\begin{figure}
[ptb]
\includegraphics[width=3.5in]{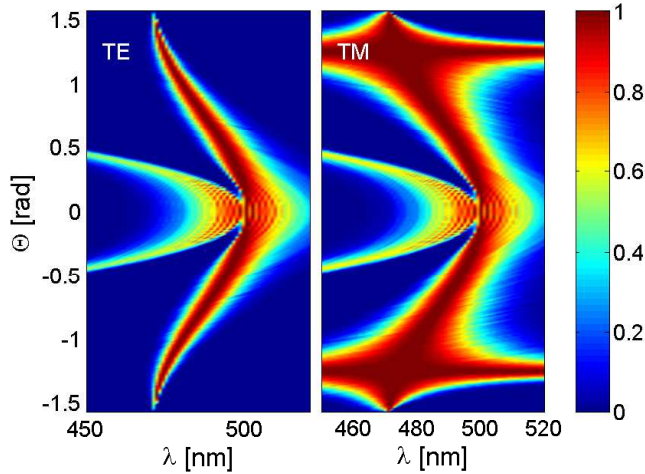}\\
\caption{The wavelength and incidence angle dependence of the transmission through a random
multilayer on a linear color scale for TE waves (left) and TM waves (right).}%
\label{3D}%
\end{figure}

Quite remarkably, it is now possible to tune the wavelength of the resonance, simply by tilting the filter.
We present the angle and wavelength dependence of the transmission in figure 3 for both the TE and TM waves using 200 random layers averaged over 100 configurations. Two branches can be observed, one which is perfectly transmitting and a second one which vanishes with increased angle. In addition, there is an angle $\Theta\simeq\pm1.25$ with perfect transmission for the TM incidence, where $\Theta=\arctan (n_1/n_0)$ rad and
$n_1/n_0=3$ for our particular configuration. This angle corresponds to the Brewster angle.

\begin{figure}
[ptb]
\includegraphics[width=3.2in]{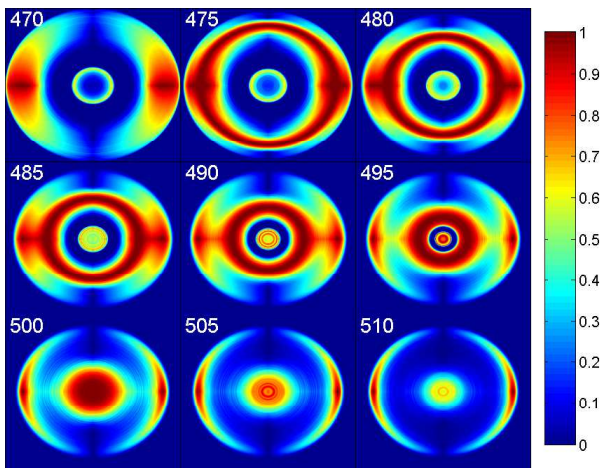}\\
\caption{The transmission is shown on a linear color scale as a
function of the angle of incidence in the $\hat{y}$ and $\hat{z}$
directions. Each subgraph spans -$\pi/2$ to +$\pi/2$ and
corresponds to a different wavelength. The resonant wavelength
($\lambda_0=500nm$) corresponds to the bottom left panel. The horizontal
(vertical) axis would correspond to the angle dependence of a TM (TE) wave.}%
\label{3D}%
\end{figure}

By combining the expression for TE and TM waves it is possible
to obtain the results for any polarization, simply by using a linear combination of TE and TM waves. Moreover, we can analyze what happens when tilting the filter in one or the other direction. The tilt angle can
be varied in two spatial directions, one corresponding to the TM direction and the other to the
TE direction, or a combination of both, which is seen in figure 4.
Perfect transmission is now identified in a form of a ring, whose diameter depends on the wavelength.
This angular dependence therefore allows for a direct and beautiful visualization of the localization-delocalization transition and hence of localization.

Summarizing, we have shown a new filter design, where we can
directly visualize Anderson localization by using a
localization-delocalization transition. The filter design is {\em
perfect} in the sense that only one visible wavelength is
transmitted perfectly and all others are exponentially suppressed.
Not only does this provide for an optimal filter design but also
allows us to {\em see} Anderson localization 50 years after its
discovery.


\end{document}